\documentclass[preprint,amsmath,amssymb]{revtex4}  
\usepackage{graphicx}
\usepackage{dcolumn}
\usepackage{bm}
\usepackage{pstricks}
\usepackage{epsfig}
\usepackage{pst-grad}
\usepackage{pst-plot}
\begin{document}
\title{Comment on ``On the general relativistic framework of the Sagnac effect" EPJC 79:187}
\author{${\rm H.\;Ramezani}$-${\rm Aval}$\footnote{Electronic
address:~ramezani@gonabad.ac.ir}} \affiliation{ Department of
Physics, University of Gonabad, Gonabad, Iran}
\begin{abstract}
In a recent paper EPJC 79:187 the general relativistic framework of
the Sagnac effect was investigated. We have some comments on this
paper. We show that their conclusion about the apparent variation of
the speed of light does not hold in stationary spacetimes. Also, We
show that their definition of gravitational Coriolis potential and
gravitational Coriolis time dilation are inconsistent with what is
stated in standard references.
\end{abstract}
\maketitle
\textbf{Comment1:} In section 4 of \cite{Benedetto} they assume a
radial ray of light to show that by definition $v=dr/d\tau$ the
velocity of light in rotating frame is always c. Here we follow
Landau in \cite{Landau} and assume a null geodesic to obtain the
velocity of light in general stationary spacetime by this
definition. The 1 + 3 (threading) formulation of spacetime
decomposition (in which the spatial line element is defined
physically through sending and receiving light signals between
infinitesimally separated observer worldlines) starts from the
following general form for the spacetime metric \cite{Landau}
\begin{eqnarray}\label{1}
ds^2 = d\tau_{syn}^2 - dl^2 = g_{00}(dx^0 - {g_\alpha}
dx^\alpha)^2-\gamma_{\alpha\beta} dx^\alpha dx^\beta~~~;~~~
\alpha,\beta = 1,2,3
\end{eqnarray}
in which $d\tau_{syn}=\sqrt{g_{00}}(dx^0 - {g_\alpha} dx^\alpha) $
is the synchronized proper time,
${g_\alpha}=-\frac{g_{0\alpha}}{g_{00}}$ is the so-called
gravitomagnetic potential and
\begin{eqnarray}\label{2}
dl^2 = \gamma_{\alpha\beta} dx^\alpha dx^\beta = (-g_{\alpha\beta} +
\frac{g_{0\alpha}g_{0\beta}}{g_{00}})dx^\alpha dx^\beta
\end{eqnarray}
is the spatial line element of the 3-space in terms of its
three-dimensional spatial metric $\gamma_{\alpha\beta}$. So we have:
\begin{eqnarray}\label{3}
ds^2 = g_{00}(dx^0 - g_\alpha dx^\alpha)^2-dl^2
\end{eqnarray}
and the velocity was introduced
\begin{eqnarray}\label{4}
v=\frac{dl}{d\tau}=\frac{c dl}{{\sqrt{g_{00}}dx^0}}
\end{eqnarray}
for a light signal $ds^2=0$ and according to (\ref{3}) we have
\begin{eqnarray}\label{5}
dl=\sqrt{g_{00}}(dx^0-g_\alpha dx^\alpha)
\end{eqnarray}
and so using (\ref{4}) and (\ref{5})
\begin{eqnarray}\label{}
v=\frac{c(dx^0-g_\alpha dx^\alpha)}{dx^0}.
\end{eqnarray}
In a static field (Such as (21) in Epjc 79:187 for radial motion) $g_\alpha=0$ and so $v=c$, but obviously in general stationary fields the velocity of light is not always c by this definition. \\
\textbf{Comment2:} According to section 4, they express in
conclusions that ``as it happens for example in Rindler or
Schwarzschild metric, the apparent variation of the speed of light
is a consequence of time dilation". It is necessary to note that
Schwarzschild metric is static while the Rindler metric and the
rotating metric (which is the subject of discussion)are stationary.
In section 4 they assume a  radial motion and so the metric becomes
static. But as we showed in comment1, their conclusion does not hold
for stationary fields. In other words, If we replace radial ray by
azimuthal ray, the apparent variation of the speed of light does not
disappear by their consideration. An important difference between
the static and stationary spacetimes is related to synchronization
of clocks over all space\cite{Landau, Rindler}. It seems that the
anisotropy in the speed of light in non-inertial reference frames is
a nonlocal effect and is related to synchronization in stationary
spacetimes \cite{Sorge,Dieks} and does not relate only to time
dilation. The velocity of light, by definition, is always equal to
c, if the times are synchronized along the given closed curve and if
at each point we use the proper time \cite{Landau}. In other words, although the speed of light is constant in all inertial reference frames and the local velocity of light is always c, but average velocities of light (which are not necessarily c) are needed for a complete description of the propagation of light in non-inertial reference frames\cite{Petkov}.\\
\textbf{comment3:} In section 5 they introduce $V=v\omega r$ as
gravitational potential corresponds to Coriolis force and then
relate the Sagnac effect to this potential. The first point is that
as we know from standard references such as \cite{Rindler}, for weak
stationary fields we have
\begin{eqnarray}\label{}
ds^2\approx(1+\frac{2\Phi}{c^2}-\frac{2\textbf{w}.\textbf{v}}{c^3})c^2
dt^2 -dl^2
\end{eqnarray}
where $\Phi$ and $\textbf{w}$ are scalar potential and vector
potential respectively. Only in the case of static metric we can
transform $\textbf{w}$ away because there will exist an $f$ such
that $\textbf{w}=-grad~f$, which happens whenever $\nabla\times
\textbf{w}=0$ and clearly not in the case of
Langevin-Landau-Lifschitz metric (equation 8 in EPJC 79:187). It is
not clear how they can obtain the Coriolis force from
the potential $V=v\omega r$.\\
\textbf{ comment4:} The equation 15 can not be the spacetime metric for eccentrically rotating observer (or as mentioned in paper: the observer on the beam splitter). See \cite{Nouri1,Nouri2} for more details.\\



\begin{thebibliography}{30}
\bibitem {Benedetto} E. Benedetto, F. Feleppa, I. Licata, H. Moradpour, Ch. Corda, Eur. Phys. J. C 79, 187 (2019)
\bibitem {Landau} L. D. Landau and E. M. Lifshitz, {\it The classical theory of fields}, 4th edn.(Pergamon Press, Oxford, 1975), Chap. 10
\bibitem {Rindler} W. Rindler, {\it Relativity: Special, General, and Cosmological}, 2nd edn.(Oxford University Press, 2006), Chap. 9
\bibitem {Sorge} F. Sorge in {\it Relativity in Rotating Frames: Relativistic Physics in Rotating Reference Frames}, ed by G. Rizzi and M. L. Ruggiero. (Kluwer academic publishers, 2004)
\bibitem {Dieks} D. Dieks in {\it Relativity in Rotating Frames: Relativistic Physics in Rotating Reference Frames}, ed by G. Rizzi and M. L. Ruggiero. (Kluwer academic publishers, 2004)
\bibitem {Petkov} V. Petkov, arXiv:gr-qc/9909081. Chap. 7 of {\it Relativity and the Nature of Spacetime}, 2nd edn. (Springer, 2009)
\bibitem{Nouri1} M. Nouri-Zonoz, H. Ramezani-Aval and R. Gharechahi, Eur. Phys. J. C 74, 3098 (2014)
\bibitem{Nouri2} M. Nouri-Zonoz and H. Ramezani-Aval, Eur. Phys. J. C 74, 3128 (2014)
\end{thebibliography}
\end{document}